\documentclass[%
 aip,
 amsmath,amssymb,
reprint,%
]{revtex4-1}

\usepackage{array}
\usepackage{graphicx}
\usepackage{dcolumn}

\usepackage{bm}

\usepackage[utf8]{inputenc}
\usepackage[T1]{fontenc}
\usepackage{mathptmx}
\usepackage{etoolbox}
\usepackage{siunitx}
\usepackage{listings}
\usepackage{tabularx}
\usepackage{multirow}
\usepackage{booktabs}
\usepackage{textcomp}
\newcommand{\codechar}[1]{\textquotesingle#1\textquotesingle}
\makeatletter
\def\@email#1#2{%
 \endgroup
 \patchcmd{\titleblock@produce}
  {\frontmatter@RRAPformat}
  {\frontmatter@RRAPformat{\produce@RRAP{*#1\href{mailto:#2}{#2}}}\frontmatter@RRAPformat}
  {}{}
}%
\makeatother
\begin{document}

\preprint{AIP/123-QED}

\title[]{A Customizable Modular Control System for Ultracold Experiments}
\author{Kaiyue Wang}
\author{Colin V. Parker}%
 \email{cparker@gatech.edu}

\affiliation{ 
School of Physics, Georgia Institute of Technology
}%

\date{\today}

\begin{abstract}
We implemented a control system for ultracold atom experiments. The system includes hardware modules that generate  synchronized experiment signals of different kinds, and a  protocol to communicate with all the modules. We also implemented software that can automatically generate experiment sequences from declarative tables of parameters with variations. Both the hardware and the software are open-source for adaptation and customization in other experiment platforms. 
\end{abstract}

\maketitle

\section{\label{sec:intro}Introduction}
Ultracold atom experiments are usually performed in units of experiment cycles, during which different kinds of signals must be generated synchronously with microsecond timing accuracy, over a period of several tens seconds, consisting of controllable process for loading and cooling \cite{dieckmann1998two, kerman2000beyond, ketterle1996evaporative}, transport and state preparation \cite{anderegg2019optical, brown2019gray, browaeys2020many}, dynamical evolution \cite{mazurenko2017cold, zhang2020pattern} and data read-out \cite{asteria2021quantum, qian2021super, bakr2009quantum}. The signals include several types: 1. Radio-frequency signals (RF) in the range of 1 to \SI{200}{MHz}, to drive acoustic-optic modulators or deflectors (AOM, AOD), to control light sources, or to drive transitions between atomic sublevels directly. 2. Analog signals to control laser frequencies through diode currents or piezo voltages, to specify currents through magnet coils, or to serve as the setpoint for proportional-integral-derivative (PID) servo circuits. 3. Digital signals to time and trigger the other events in the experimental sequences, such as data taking, and to open and close mechanical shutters. The capability to generate and customize these signals can provided through either commercial modules or integrated circuits, both methods requiring work to customize according to the need of the experiment and integration into the existing systems and users. In the past decade, we have seen an increasing number of open-sourced AMO control system solutions in both hardware \cite{shammah2024open, artiq} and software \cite{starkey2013scripted, perego2018scalable, keshet2013distributed}. Here we share a set of designs encompassing a complete system, from the schematics of each signal module to the overall communication and control over these modules from the terminal.

\section{\label{subsec:modules} Signal Modules}
In our setup, each module usually has multiple output channels. An experiment sequence is composed of partitions for every output channel, each described by a table of multiple parameters at each frame, which we call the ``parameter table''. The frame rate of the experiment is decided by a common clock signal, in our case \(f_\text{Rack CLK}=\SI{50}{kHz}\). Therefore, the full parameter table has a row number of \(\Delta t\cdot f_\text{Rack CLK}\). These parameter values (the columns of the table) are of different types for different types of modules. When a sequence runs to a certain frame, a module can either be idle if all it's output channels maintains the values as the previous frame, or perform an update if any of the channels should change it's value. We call the latter frame an ``update-frame'' for the module.

\begin{table*}
\label{tab:modtype}
\caption{Selected list of integrated module types and their corresponding output signal, channel number, tunable parameters and ranges as currently implemented.}
\begin{tabularx}{\linewidth}{XXXXXX} \toprule
\textbf{Signal}                           & \textbf{Chip}                    & \(N_\text{chan}\) & \textbf{Tunable Parameter}   & \textbf{Range}    &\textbf{Ramp}      \\
\midrule
Digital& TM4C123GH6PM & 16 & binary value & 0 or \SI{3.3}{V} & No \\ \midrule
\multirow{3}{*}{Radio-frequency} & \multirow{3}{*}{AD9959} & \multirow{3}{*}{4} & amplitude           & 0\textendash\SI{100}{mV}(RMS)   & Linear    \\ 
                                 &                         &                    & frequency           & 0\textendash\SI{200}{MHz}  & Linear    \\
                                 &                         &                    & dithering amplitude & 0\textendash\SI{100}{\%} & Linear      \\ \midrule
Analog                           & AD5686 AD5750-2         & 4                  & amplitude           & \SI{-10}{V}\textendash\SI{10}{V}  & Linear    \\ 
\bottomrule
\end{tabularx}
\end{table*}
Each signal module is designed to generate a specific type of signal to multiple output channels. A module usually hosts a micro-controller unit (MCU) that takes care of the logic and one or more signal generating chips that generates the signal (See \ref{tab:modtype} for a selection of module types). One exception is the digital module, where the signal source is the General Purpose Input Output (GPIO) on the MCU itself. Before an experiment cycle runs, the MCU stores a compressed parameter table for each of the channels it controls. When the sequence starts, the MCU is notified by a start trigger and on each clock cycle, it checks whether the frame is an idle frame or update-frame. For each update-frame, it generates commands to it's associated chip (other than digital) through Sychronous Serial Interface (SSI, also known as Synchronous Peripheral Interface) lines and updates the output.

\begin{figure*}
    \centering
    \includegraphics[width=\linewidth]{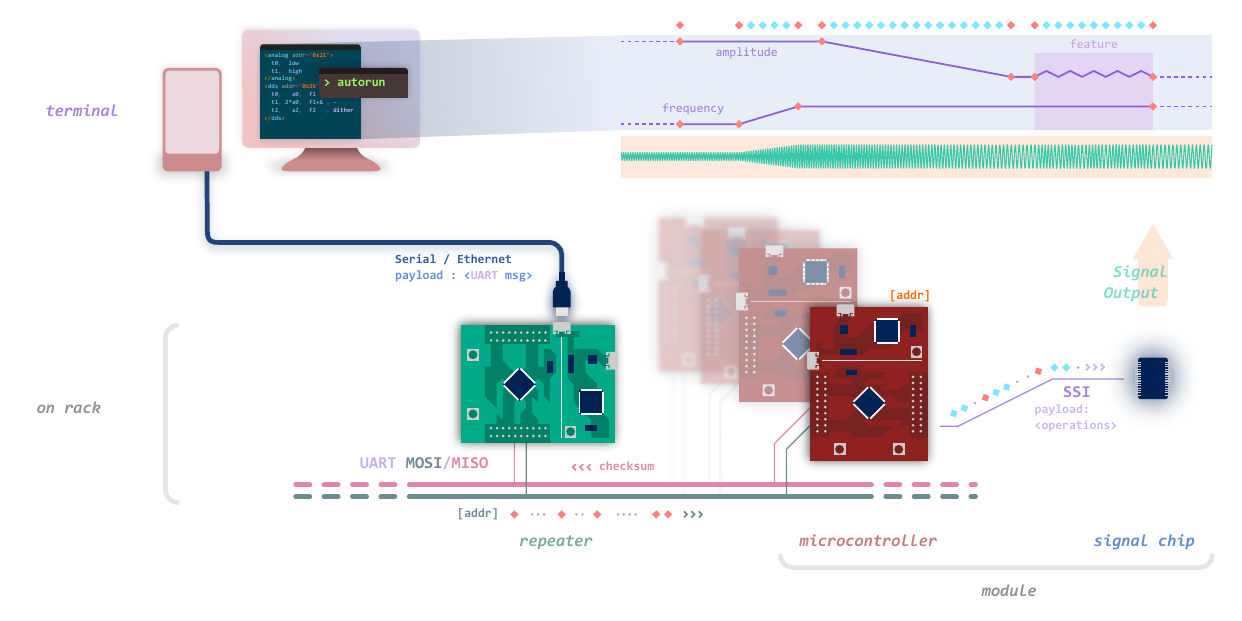}
    \caption
{The information flow when loading a sequence, from terminal to the repeater, to the MCU, to the signal chip. The sequence is being decompressed at the chip.}
    \label{fig:comm}
\end{figure*}

Since the entire parameter table for each module is very large in size if the sequence contains ramps, the table is compressed into a list of commands before being sent to the module, containing only checkpoint data to save the bandwidth of communication. The compression and decompression is elaborated in \ref{sec:runcycl}.

\paragraph{Digital Module}
The \num{16}-channel digital module directly uses the GPIO port on the MCU (EK-TM4C123GXL), so it does not need an associated chip. Instead, each channel output goes through one of the two SN74HC377N D-Type Flip-Flop acting as a buffer to synchronize with the Rack CLK. The signals are output from the front panel through two 9-pin D-Subminiature (DE-9) in either \num{0} or \SI{3.3}{V} and they are updated per frame.
\paragraph{Radio-frequency Module}
The RF modules uses one quad-channel AD9959 chip to generate RF signals on \num{4} output channels, each with per-frame control on the RF amplitude and frequency and has a resolution of \SI{10}{Bits}. We implement the dithering feature utilizing the chip's ``modulation mode''. The dithering sweeps the RF frequency within a custom range, which can be used for, e.g., expanding beam width after an AOM. In the current implementation, the dithering waveform is controlled by fixed inputs generated from the MCU, while the frequency sweep range (the dithering amplitude) can be updated per-frame. The module outputs signal through the \num{4} SMA ports on the front panel. The AD9959 relies on a \SI{50}{MHz} reference clock which relies on another SMA input on the front panel.
\paragraph{Analog Module}
The analog module uses one AD5686 quad channel 16-bit Digital-to-Analog Converter (DAC) chip to general analog signals on 4 output channels, each channel's voltage output (within range \SI{0\textrm{--}2.5}{V}) controls a AD5750-2 output driver to adjust to different voltage ranges and provide higher output currents. The supported output ranges are \SI{0\textrm{--}5}{V}, \SI{0 \textrm{--}10}{V}, \SI{-5\textrm{--}5}{V}, \SI{-10\textrm{--}10}{V} and can be configured per channel for each cycle run, while the DAC output can be updated per frame. The signals are output from the front panel through four BNC ports. 

\section{\label{sec:com}Communication Protocol}
\subsection{\label{subsec:ssi} SSI within a Module}
The communication between the MCU to a signal generating chip usually relies on the SSI according to the chip's protocol. Values of an update-frame targeted at different output channels or parameter types are distinguished by the different registers in the SSI message. The transmission is synchronized by a \SI
{20}{MHz} clock generated by the MCU, while the output for a chip should be able to update up to every \SI{20}{\micro\second}, limiting the amount of information sent within one frame to less than 400 bits. For each module type the number of signal updating operations we can write to the associated chip is fixed, which we refer to as the ``slot''. The slot number especially restricts the DDS modules and we must avoid updating more than 4 values in one frame. The MCUs can be found with higher SSI clock rate support and/or multi-wire transmission formats, so this restriction could theoretically be relaxed, but we find that in practice this is adequate.

\subsection{\label{subsec:comsync} Repeater, COM Bus}
The computer terminal cannot be connected to all modules at once for communication. In other work, the update commands are usually carried out by a master FPGA module through multi-bit bus line during the run cycle \cite{perego2018scalable, bertoldi2020control, trenkwalder2021flexible, keshet2013distributed}, and multiple master modules needs to be used for extended modules. In our system, since each module is associated with a MCU, we only need to setup a special module, which we refer to as the ``repeater'', to pass information between the terminal and each module. The repeater can communicate with the terminal by a computer-to-repeater UART link (e.g. using a virtual COM port through it's USB connection, or TCP via an Ethernet cable), and it communicates with all of the other modules before execution via by four common bus lines that we call the ``COM Bus'': 

\begin{table}
    \centering
    \begin{tabularx}{\columnwidth}{cXX} \toprule
\textbf{Name}      & \textbf{Description}                                               & \textbf{Direction}                           \\
\midrule
UART MOSI & Update parameters for all modules                         & Repeater to regular modules               \\ \midrule
UART MISO & Checksum                                                  & Regular modules to repeater               \\ \midrule
Rack CLK  & Experiment synchronization clock signal (50 kHz)          & from the clock module to others           \\ \midrule
START     & The experiment ``Start'' and ``Abort'' signal for all modules & from one leading digital module to others \\ \bottomrule
\end{tabularx}
    \caption{The 4 lines in the COM Bus.}
    \label{tab:com-bus}
\end{table}

When editing the parameter tables of each module, the repeater broadcasts the module-specific command through UART MOSI (Master Output Slave Input) to each regular module sequentially and checks for a response with checksum on UART MISO (Master Input Slave Output). This UART broadcasting protocol allows new modules to be installed into the system, and the interface can be easily implemented through the MCU's peripheral modules in both hardware connection and software handlings.

\subsection{\label{subsec:uart}UART data Transfer}
By the UART protocol, data are formatted into frames. In our case, each frame contains 8 bits, or 1 byte of actual data. A transmission line is held high in idle. Each frame of data consists of a start bit, 8 data bits, the 9th bit as address indicator, and one stop bit. See Fig. \ref{fig:parse-load} \textbf{b}.

Since there are only two UART lines connecting every module: MOSI and MISO, multiple modules should not be responding at the same time, and typically the broadcast message is targeted to only one module. In our case, the 9-bit address mode is configured in the regular modules as the receiver. Each regular module will actively listen to all the signals on the UART MOSI line, but will only accept data when the receiver is in the address matched state. An address match occurs if the following conditions are both true: first the frame's 9th bit is set, and second the frame's 8-bit data is a byte that matches the device's programmed address. After matching the address, the device receives as data the first 8 bits from any additional words that have their 9th bit clear, until another address message with a non-matching address is sent, at which point data words are ignored and the receiver exits the address match state.  Since all modules have different UART addresses, only one module receives the message at a time. This addressing mode is supported at the hardware level. The 9th bit is essentially an out-of-band signal that prevents confusion when a message to one module may contain a data byte that matches the address and message start structure of another module, which would accidentally start data transmission to other devices. Additionally, transmission to one module does not affect the instruction sequencing of the other modules, since no interrupts are generated by the hardware without an address match.

In general, we formalize the message in the following way: 1. A single byte \texttt{0xAA} indicating the start of the message, where the alternating bits serve as a check for noise triggered false messages. 2. The aforementioned command list containing the compressed sequence for that module. The command list has an indefinite length, but always ends with a command prefixed by \texttt{\codechar{T}}.

\subsection{\label{subsec:dig}Clock, Start, Digital, Repeater All in One}
Compared to other module types, the digital module contains some unique elements. For one, it doesn't have a distinct signal generating chip, but uses the MCU itself to output TTL digital signals. It can also set the START signal in ``Leader Mode'' and the \SI{50}{kHz} synchronization signal in ``Clock Mode'' and broadcast them to the respective lines in ``Repeater Mode''. In the latest implementation, it can also be connected to the terminal through the USB port on the evaluation board that hosts the MCU, and act as the repeater. There could be multiple digital modules connected to the system, but only one of them is set to the Leader/Clock/Repeater Mode that enables the aforementioned functionalities (and they could be the same one). For the module in Repeater Mode, it uses two different UART ports: A virtual COM port is used to receive the messages from the terminal, and the connection to the MOSI line broadcasts the messages to others. When the address matches itself, it directs the messages internally without going through the COM Bus. It is a repeater and a regular module in one.

\section{\label{sec:runcycl}Typical Run Cycle}

\subsection{\label{subsec:cmdlist}Parsing Stage, Command List}
For an experimental cycle involving multiple modules, the sequence is first edited on the terminal computer. A program (currently implemented in MATLAB) parses the editable sequence file and the parameters to multiple command lists, one for each module. The command list follows some common protocols. Each command starts with a leading prefix character indicating the command's type. The body of the command depends on the prefix and the module type. 

\begin{figure}
    \centering    
    \includegraphics[width=\columnwidth]{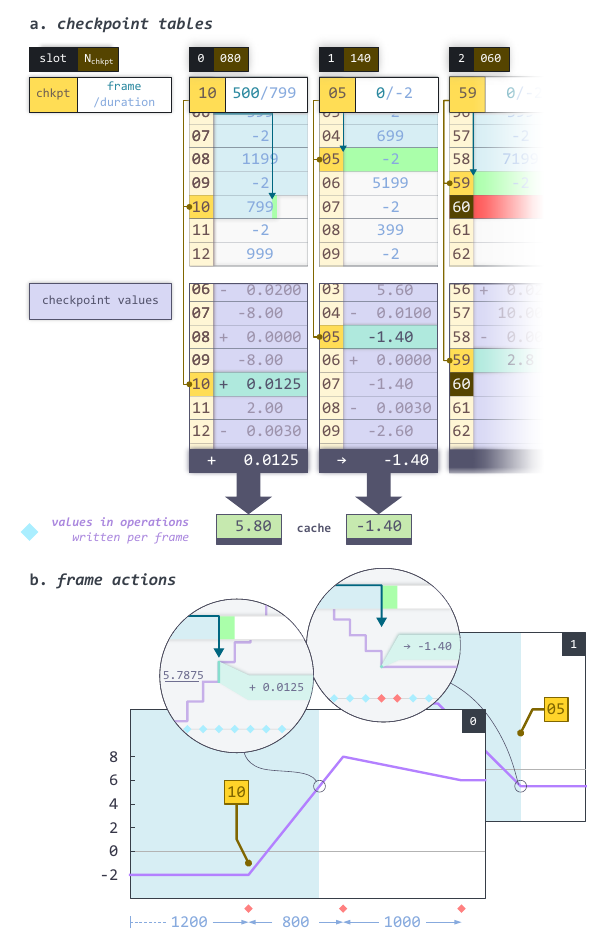}
    \caption{\label{fig:exec} On-the-fly MCU logic. \textbf{a} Each checkpoint table is labeled by the slot index. On every frame the cache value is updated on each slot by reading the duration and value table entries of the latest checkpoint index. The updated value is then written to the chip through an operation. In slot \num{0}, the sequence has completed \num{500} frames of a ramp whose length is \num{799} frames. To generate the \num{501}\textsuperscript{st} frame, the cache value, \num{5.7875}, is updated by adding the increment value, \num{+0.0125}, to reach \num{5.8}. In slot \num{1}, the sequence just finished a ramp with length \num{699}, and proceeds to a ``jump'' checkpoint marked by a duration of \num{-2} but lasts for one frame, which sets the cache value to the absolute value \num{-1.4}, as recorded in the corresponding value table entry. Slot \num{3} demonstrates the last checkpoint is at position \num{60}, as determined by \(N_\text{chkpt}\) in the \texttt{\codechar{E}} commands in Fig. \ref{fig:comm} \textbf{c}. \textbf{b} The effect of the operation on this update frame, causing the output signal to update. In this example, the two slots correspond to two different output channels, but they can also be two parameters of the same channel. }    
\end{figure}

The checkpoints' data is mainly represented by the ``update commands'' with prefix \texttt{\codechar{U}}, whose body includes slot, duration, parameter values. A ``ramp'' checkpoint carries a positive duration, and the associated values are interpreted as increment on the target, and the duration is the number of frames for that value increment to take effect on each frame. For all modules, the ``jump'' checkpoint uses a negative duration to indicate that the associated value is absolute, which can rectify rounding errors (See Fig. \ref{fig:exec}). For modules that support linear ramping (See \ref{tab:modtype}), update commands are mostly with alternating-signed duration. These data are directly used in the execution stage \ref{subsec:exec}, saving the MCU from more expensive calculations such as divisions.

The update command's remaining length depend on the module type, and can contain a mask for targeted channel and parameter type. Other prefixed commands have a more uniform structure across module types, containing instructions on formatting the MCU's memories, and the start and abort of the execution (See Fig. \ref{fig:parse-load} \textbf{c}).

\subsection{\label{subsec:load} Loading Stage}
In the data loading stage, the terminal sends each command list along with the target module's address to the repeater by the USB-COM connection, who then repeats the message to the target module on the UART MOSI line, and waits for the response of that module on the MISO line before sending additional messages to other modules. Within this time, the terminal will not hasten to send command lists of the next module. Even if does, it would be blocked by the repeater and be warned of its busy state. This removes the possibility of multiple modules having conflicting responses on that line. The response of the module is usually a checksum. If it matches the expectation of the receiver, or the response time exceeds timeout, the repeater will report this back to the terminal. Only after a report is received from the repeater will the terminal program proceed to the next command lists, and the process will loop until all command lists are sent. It is not common, but a checksum error could occur for some of the modules, which most likely results from a bit error, or a false initial state from previous interruptions. Both of which are usually resolved by re-sending the messages from the console. But if all things go well, this stage typically takes \(2\text{--} 5 \,\si{sec}\). In applications where only small changes are made between experiments, this time could be reduced substantially if only the updated values are rewritten.

\begin{figure*}
    \centering
    \includegraphics[width=\linewidth]{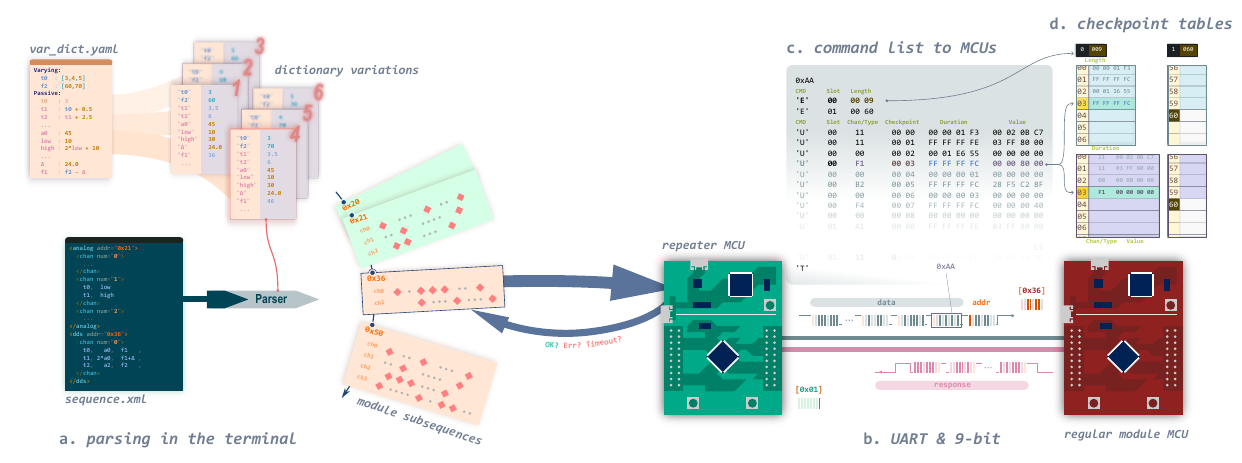}
    \caption{\label{fig:parse-load}Parsing and the run cycle's updating stage for one of the modules. \textbf{a} The terminal program generates command lists (c) by parsing the \texttt{yaml} variable dictionary and the \texttt{xml} sequence file per module. \textbf{b} On the UART line, the 9-bit address is reflected by the trailing bit. \textbf{c,d}  Each line in the command list acts on the checkpoint tables, which are maintained in the MCU's memory. The most two common commands \texttt{\codechar{E}} and \texttt{\codechar{U}} extend the table length and update the table entries, while the command list ends with \texttt{\codechar{T}}.}
\end{figure*}

The logic of this stage is implemented by a UART-handler function in the MCU program that builds a local checkpoint data table as per the command list. The handler would be called only if the address check passes. For each module, it uses the command list to build internal checkpoint tables that contains checkpoint data as shown in Fig. \ref{fig:parse-load} \textbf{c}. The tables (Fig. \ref{fig:exec}) are indexed by the slots and checkpoint indices.

\subsection{\label{subsec:exec}Execution Stage}
After we confirm that all modules that are meant to be updated have loaded their parameter tables, a start command is sent from the terminal as the start of the execution stage. The repeater will notify the digital module in Leader Mode with an \texttt{\codechar{S}} command with no trailing data, who will set the ``START'' line to high. The rising edge notifies every modules to start their execution simultaneously. 

By each ``Rack CLK'' cycle, the frame position increments by one. The MCU finds the corresponding actions at the current frame by referring to the checkpoint table by the rules in \ref{subsec:cmdlist}, including incrementing or jumping to parameter values. A ramp-supporting module maintains a set of cache value, one for each channel and parameter to keep track of the latest parameters. On the update-frame the program reads the cached value, and perform an operations per slot, which includes multiple commands to the chips through SSI. 

Fig. \ref{fig:exec} shows the table entries as floating-point values, but the actual tables always store 32-bit integers (except for the digital values that uses one bit for each channel). For a ``ramp'' checkpoint, the entry values are interpreted as signed integers (\texttt{0xFF} is \num{-1}), enabling ramping in either directions, while for a ``jump'' checkpoint, the entry contains absolute values that act like unsigned integers (\texttt{0xFF} is the maximum value), even if the values may represent output signals smaller than \num{0}. A chip operation usually uses data less than 32 bits as specified by the protocol, so only the higher bits are used in each operation (truncated by bit-shift) but the lower bits still help the accuracy through the cache values so that less error accumulates. When the table values are generated in the command list from the terminal, they are adjusted with a small offset to account for the rounding errors by truncation.

\section{\label{sec:hwencap}Hardware and Encapsulation}

\begin{figure*}[t]
    \centering
    \includegraphics[width=\linewidth]{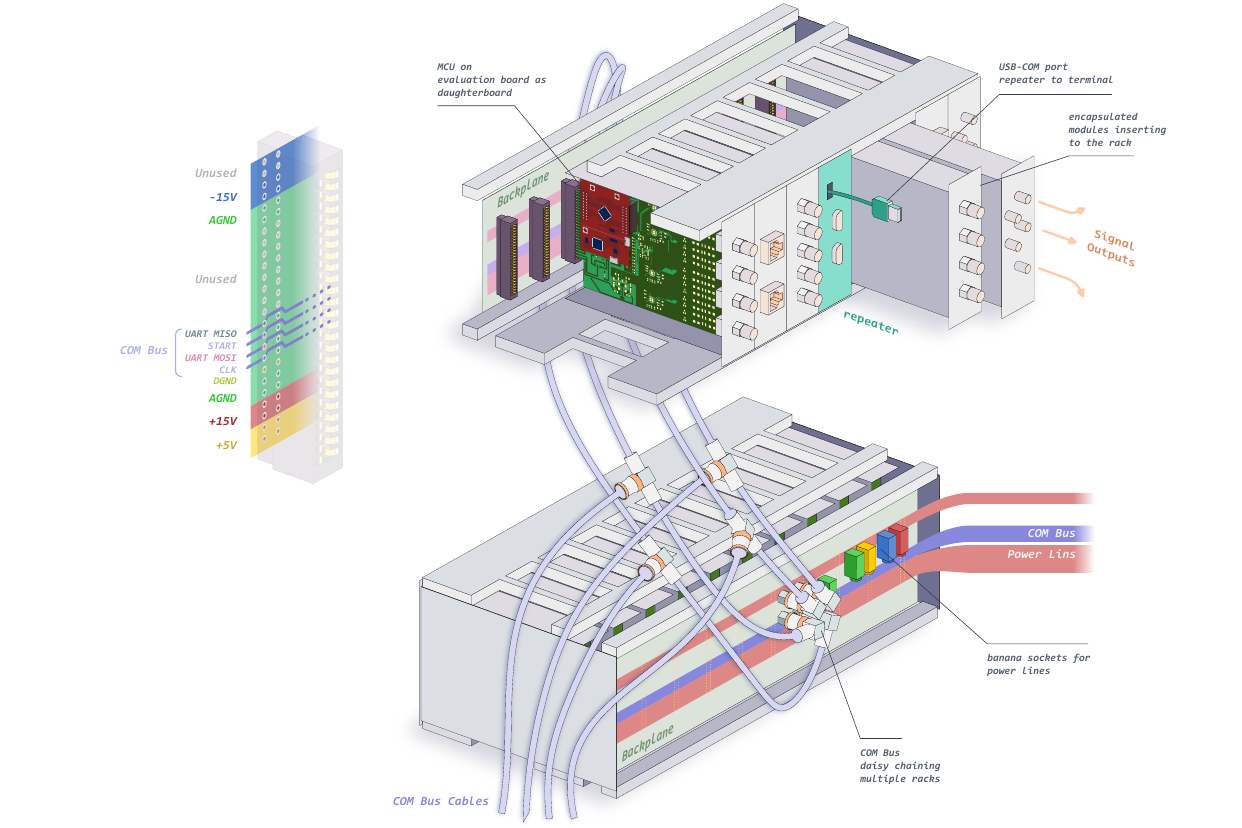}
    \caption{\textbf{a} The in configuration for the card edge connectors. \textbf{b} The front view on the rack that hosts multiple modules encapsulated in cases. The multi-channel signal outputs are from the front, and card edge connections are to the back. The USB connection for the repeater is also extended from the front panel. \textbf{c} COM Bus are connected by cables with BNC connections to daisy chain multiple racks. Power lines are connected through banana sockets.}
    \label{fig:hw-conn}
\end{figure*}

To ensure a consistent performance among all modules of the same type, each module is based on a Printed Circuit Board (PCB), which integrates the physical input-output (I/O) ports, connection to power supplies and 4 COM Bus lines, connection and mounts to the MCU, and some signal processing components. 

Each PCB has a size of \(\SI{4.5}{in} \times \SI{6.5}{in}\). On one end of the the boards (which we call the ``back''), they adopt the same edge connector, consisting 25 rows of exposed traces on each side, among which 11 rows are used as power supplies (\SI{5}{V} and DGND for TTL power supply, \(\pm\SI{15}{V}\) and AGND for analog power supply, these rows are connected to both sides), 4 rows for COM Bus, 8 rows are for customizable data bus and the 2 rows on the top are disconnected (See Fig. \ref{fig:hw-conn} \textbf{a}). On the other end (front), there are fixed positions for mounting I/O connectors. They are usually BNC/SMA connectors as signal outputs channels, diagnostic monitor channels or input channels like triggers and feedback signals or reset switches.

For the MCU, we employ the Texas Instrument evaluation module EK-TM4C123GXL, which in fact hosts two identical MCUs (TM4C123GXL) and peripheral features such as UART and Pulse-Width Modulator (PWM). Each module's PCB is designed with two double rows of 0.1'' pitch headers, which allows the MCU evaluation board to be mounted through it's bottom header expansions as a daughter board (See Fig. \ref{fig:hw-conn} \textbf{b}), meanwhile powering the MCU evaluation board and routing the communication between the MCU and the rest of the module. 

A PCB along with its mounted components and daughter board can be hosted in an aluminum case (EFP164A66), which can be inserted to aluminum module racks (CMA13-16/90). Each rack can hold up to 10 modules of this size. We modify the front panels of each module to match the I/O connectors' positions on the PCB, so that the physical connections can be configured without opening the case. All modules are connected to a common ``Backplane'' mounted on the back of the rack through the card edge connections. The Backplane is also a PCB of dimension \(\SI{16}{in}\times\SI
{4.5}{in}\)
. It mounts 10 card edge sockets aligned with the position of the modules, where the socket pins are connected through the Backplane PCB traces. For the power lines, it uses \SI{5}{mm} banana jacks as input from the DC power supply, and one push switch to power on/off the rack and its hosted modules with relays. The card edge socket of the same power row are connected by the traces on the Backplane. For the COM Bus lines, it uses the mounted BNC connectors, where the shields are connected to the ground side of the card edge sockets. Only one side on the card edge socket are connected to the corresponding COM Bus lines, where the other side pins are grounded for shielding. 

Since the MISO line has multiple modules as candidate drivers, they cannot be active at the same time, while simultaneous disconnection would leave the line floating and is prone to noise signals and false responses. Therefore, the MISO line is integrated with a \SI{50}{Ohm} pull-down resistor to the ground on each Backplane. When a module is ready to respond, it will first activate the MISO line by setting it to idle state (high), and the line will be pulled down again after the message terminates. These two edges will not trigger a UART handling from the repeater.

Multiple racks can have their communication lines daisy-chained together. In this case, only one rack would host the repeater module. In our lab, we currently use 5 different racks mounting [How many] modules together, the distance between the racks that are furthest apart can reach up to \SI{10}{m}, but the commands are transmitted without errors for the greater most of the time.

\section{\label{sec:ext}Prototyping and Extensibility}
Although we have designed PCB for most of the major module types, other kind of modules can be integrated in this system. In fact, many of the prototype modules needs to be encapsulated in two cases, one hosting the MCU on a custom breadboard that has the same card edge, the other hosting the evaluation board of the signal generating chip (which is also the reference of our later PCB designs) with no card edge connection to the Backplane. Jump wires or ribbon cables are used to pair the two boards. Many of these prototype boards are still used and working in our lab to this day.

To build a newly typed module with custom functionalities, we need to match its interface to the Backplane, including COM Bus and power connections, as well as exposing physical I/O port on the front panel. If a MCU is used, its program may have a UART handler that aligns with the sequence format from the terminal to change its parameter table, and a GPIO handler to decide the action on execution frames.

The programmable MCU integrated in our scheme offers ability to do complex tasks that beyond that of an analog circuit. Notably, we have created modules that can generate \si{\micro\second} pulses, and modules that can scan and read a spectroscopy for laser locking. An example of the extensibility is to modify a PCB analog module into a frequency feedback module, which takes an RF wave, and outputs an analog voltage as the feedback control to the RF wave's frequency, in order to maintain this frequency to a constant value. This module is useful for, for example, locking a laser's frequency in the middle of the experiment, where the input frequency comes from an frequency beat with a variable reference frequency. The RF input is connected to the MCU through pairing from a neighboring module, counted by the MCU's clock counter. The feedback logic is implemented by modifying the MCU's program, which for each frame, compares the counted frequency with a pre-determined constant, and the output (or diagnostic information) replaces one of the analog output channels. Currently we implemented a logical PID feedback, with the gains as parameters that can be modified by the terminal, but it can be enhanced with a more complex logic, e.g. with a searching algorithm, so that it can attempt to re-lock itself when the lock fails.

\section{\label{sec:terminal}Sequence Generation from the Terminal}

We designed an ergonomic workflow to build, edit, vary, and document a sequence that involves multiple modules. 

The structure of the sequence is layout through an \texttt{xml} (eXtensible Markup Language) file. The parent-child hierarchy of tagged elements is fitting to describe the hierarchy of modules and output channels, so usually a document contains multiple module elements, and each module tag contains multiple output channel elements. The text body of an output channel elements contains an csv-like table, where each row is a row of the parameter table on the checkpoint, along with the time of that checkpoint. The digital sequence is again special, where all channels are fit into one module element, and the channel number makes up a column. Constant data like UART address, Rack CLK frequency, pulse width, gains, features can be encoded by named attribute of the channel or module elements. Each module element must have a UART address. What makes the scheme flexible is that the fields in the parameter table doesn't need to be numerical, but can be expressions using variables. The program that parses the \texttt{XML} sequence will traverse the DOM tree to evaluate every table, and it converts each table to numeric by evaluates the expressions with a variable dictionary, which has strings as keys and numbers as values, and is passed in to the parser when it's called.

The variable dictionary is constructed from a \texttt{YAML} file, which has key-value pairs separated by colons to represent items in the dictionary. The key-value pairs in the file aren't necessarily numeric either, but can be expressions that depend on previous pairs. Our version of the dictionary constructor program is also in MATLAB. Naively, it reads the key-value pairs sequentially, evaluating the expression value into numeric and adding it as the dictionary item. Additionally, the program allows generating multiple dictionaries at once. We define a block of \texttt{YAML} pairs under the block' key \texttt{Varying}, each of the pair in that block has a \texttt{YAML} sequence of numbers as the pair's value, and we move all other pairs under the block's key \texttt{Passive}. In this case, the constructor generates dictionaries of all possible value combinations from the \texttt{Varying} block. With a given index of variation, the constructor will first evaluate the variables in \texttt{Varying} by selecting each variation of each variable determined by that index, then proceed on construct the \texttt{
Passive} part in which the values can still depend on the \texttt{Varying} variables. Sequence values in the \texttt{Varying} are mostly numeric or ones that do not depend on other variables, while \texttt{Passive} values are mostly expressions in \texttt{String}. \texttt{YAML} is a superior choice compared to other format since it does not require quotation marks around strings by default, and it supports comments.

Typically during one experiment test, we iterate over all the variation defined by the \texttt{YAML} scheme file by the variation index. In each iteration a unique dictionary is constructed and passed to the parser to parse the same \texttt{XML} sequence file to acquire a numeric sequence. This sequence is then organized according to their target modules, and then it execute an experiment cycle as described in \ref{sec:runcycl}. During the cycle, other terminal programs are responsible for taking data from the devices connected to the terminal, and simple on-the-fly analysis. The program will also serialize and save the numeric sequence into a \texttt{json} file, along with the \texttt{XML} and \texttt{YAML} file for the record. 

Separating the \texttt{XML} sequence with variables facilitates us to re-use parameters for different stages of experiment. For example, in a cold atom preparation, after we optimize the MOT-loading stage's parameters with a sequence that ends with imaging at the MOT, these parameters are automatically applied to subsequent stages of dipole loading and evaporation cooling, so that we only need to maintain one set of parameters in the \texttt{YAML} file. Compared to GUI-based \cite{keshet2013distributed, perego2018scalable} and scipt-based \cite{starkey2013scripted} user interface, using configuration files of declarative formats clearly separates the parameters and logic, which makes it easier to edit and review. For example, we can quickly compare the current parameters with a previous day's parameter without running the script in a Matlab environment. Multiple variable tests also increases the automation in the experimental process, and organizes tests with similar purpose more cleanly.

\section{Conclusion}
We introduced the control system for an ultracold atom experiment platform in our lab from software to hardware, from circuits to encapsulation. The system satisfies our requirement for reliable, synchronized, flexible experiment signal generation and control in our ultracold atom experiments \cite{long2021spin, wang2023instability}. It is based on modules that can be extended in numbers and functionalities and many protocols can be adapted by experimental Labs and Teams in similar field and beyond. We open source our designs of all the PCB modules, the program of MCUs and terminals, as well as the test runner framework.   

%

\end{document}